\documentclass[aps,twocolumn]{revtex4}
\usepackage{graphicx}
\begin{document}

\title{\large \bf
Submarine Landslides and Local Tsunami Waves (Corinthos Gulf,
Greece)}
\author{
\normalsize $L. I. Lobkovsky^a,  G. A. Papadopoulos^b,  I. A.
Garagash^c, O. R. Kozyrev^d,\,\, and \,\, R. Kh. Mazova^d$ \\ {(a)
\it Institute of Oceanology, Russian Academy of Science, Moscow,
Russia}  \\ (b) {\it Institute of Geodynamics, National
Observatory of Athens, Athens, Greece} \\ (c){\it Institute of
Physics of Earth, Russian Academy of Science, Moscow, Russia} \\
(d) {\it Nizhny Novgorod State Technical University, Nizhny
Novgorod, Russia}}

\begin{abstract}
It is considered a problem of submarine sediment slide which
generates the surface water waves. To simulate numerically the
landslide motion it is used the method which permits to take into
account detailed rheological properties of slide body
constituents. The numerical simulation of landslide-generated
surface water waves is performed on the basis of nonlinear shallow
water equations. It is studied a landslide behavior for two values
of maximal friction angle. The detailed comparison of landslide
dynamics and evolution of landslide-generated surface water waves
during sliding is performed. Also, the evolution of dipolar water
wave generated in the beginning of sediment sliding is studied. It
is obtained that this dipolar wave is then transformed in two wave
groups: crest and trough coming seaward and trough accompanied by
crest coming to the shoreline. The second wave group leads firstly
to sea recession from the beach and only then to large runup
(tsunami wave with first negative phase).
\end{abstract}
\maketitle

Large water waves produced by submarine landslides were observed
in many regions of the world for last 50 years (local tsunamis) (
see, e.g. \cite{GLKM}). For numerical simulation of
landslide-generated tsunami there were proposed a number of
models, main of which are a rigid-body model and viscous
(visco-plastic) model (for review, see, e.g. \cite{Fine2}). First
model, because of its specifics, overestimates the water surface
response to the submarine perturbation while second model
underestimates it. To simulate adequately the landslide-induced
tsunami it is necessary to use methodics which take into account
the both detailed structure of landslide body and mechanical
characteristics of slide-body constituents during sliding (see,
e.g.\cite{MulAl}). The character of disconsolidation of
constituents in sediment surface layer appears to be a key factor
which controls the localization process of the plastic strain and
thus the slope instability. The model numerical simulation of
sliding process at the continental slope was carried out for a
number of slope parameters. In particular, such simulation of
landslide induced tsunami was performed for parameters of
Corinthos Gulf, Greece (see fig.1, 2) where in 7 February 1963 it
was observed a damaging tsunami wave formed without any seismic
events: mass of unconsolidated sediments slumped into the sea
water around of the local river \cite{KP}.
\begin{figure}
\includegraphics[width=8.0 cm]{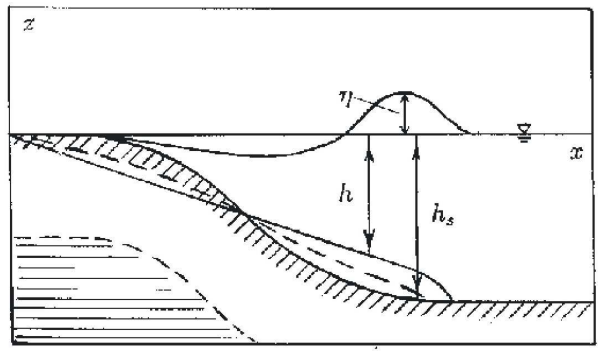}
\caption{Model scheme for simulation of tsunami wave generated by
moving submarine landslide}\label{disp1}
\end{figure}
It is considered layer-by-layer sliding of upper part of
elastically-plastic sediment layer on the slope surface which is
formed during the landslide process. It is proposed a distinct
interface between water and landslide body with water density
$\rho_w(x, z) = const$, and landslide density is a function of
coordinate $\rho_s = \rho_s(x, z)$ (see, fig.1). The coordinate
origin is taken at the coastline, with $x$-axis at the undisturbed
seawater level and directed seaward and with $z$-axis directed
upward. In the fig.1 $z = - h(x, t)$ is a variable depth of
seawater, $z = - h(x, 0) = - h_s(x,0)$ is a submarine slope
profile before initial action ($t = 0$). $D(x, t)$ is a landslide
body thickness: $h(x, t) = h_s(x, t) - D(x, t)$. The shape of
seawater surface $z = \eta(x, t), \quad \eta(x, 0) = 0$.
\begin{figure}
\includegraphics[width=8.0 cm]{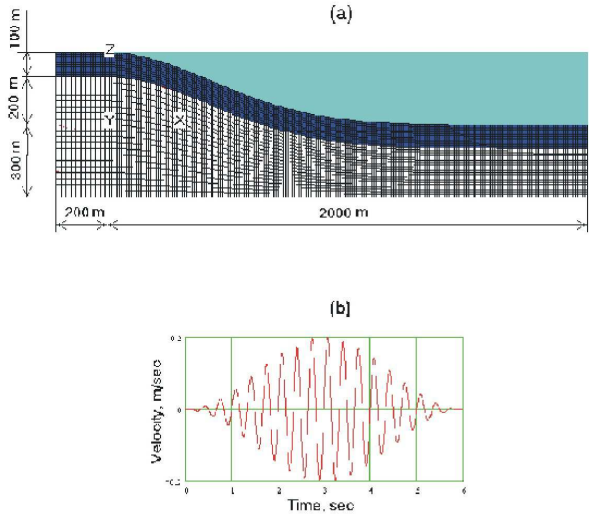}
\caption{a) Simulation scheme of submarine slope; upper layer is
an elastically-plastic sediment mass rested on elastic base; b)
Change of base velocity during earthquake with duration $6 sec$
and frequency $3 Hz$; maximal velocity is equal to $0.2 m/sec$
corresponding to earthquake magnitude equal 6} \label{disp1}
\end{figure}
At simulation it was used explicit finite-difference scheme which
permits to simulate the nonlinear behavior of pore saturated
sediments under conditions of plastic flow above yield stress.
Also, possible initial action at the process (adequate to some
seismic action) was taken into account (see fig.2, where it is
presented a geometrical scheme for numerical simulation of
landslide (fig.2a) under the action of earthquake with moderate
size magnitude (equivalent to slumping volume $57.000 m^3$ in
Corinthos Gulf), fig.2b).

It was used a layered model of sediments rested on relatively
rigid base. For each layer there were taken a layer density, shear
modulus, bulk modulus, cohesion, maximal friction angle, and
tensile strength. At the first stage, it was made the simulation
of initial, preliminary stressed state of slope with sediments
formed under action of its own weight and saturation with the
water under sea pressure. At the second stage, it was taken into
account the rheological effect of decrease sediment mass strength
above yield stress, pore saturation of sediment mass and
possibility of landslide-mass liquefaction under conditions of a
seismic or any external action. And finally, it was performed
numerical simulation of surface water waves (evolution and runup)
on the basis of nonlinear system of shallow water equations with
using of constructed explicit difference scheme with fulfilled
stability conditions. Dynamical interaction between landslide
motion and surface waves was taken into account via continuity
equation. The results are consistent with those in conventional
(rigid-body and viscous-fluid) model and agree qualitatively with
natural data. So, it was really observed the sea recession (first
negative tsunami wave) in the region of estuary of the local river
in Corinthos Gulf, Greece where coastal strip have been slumped
into the sea water (see above).
\begin{figure}
\includegraphics[width=8.0 cm]{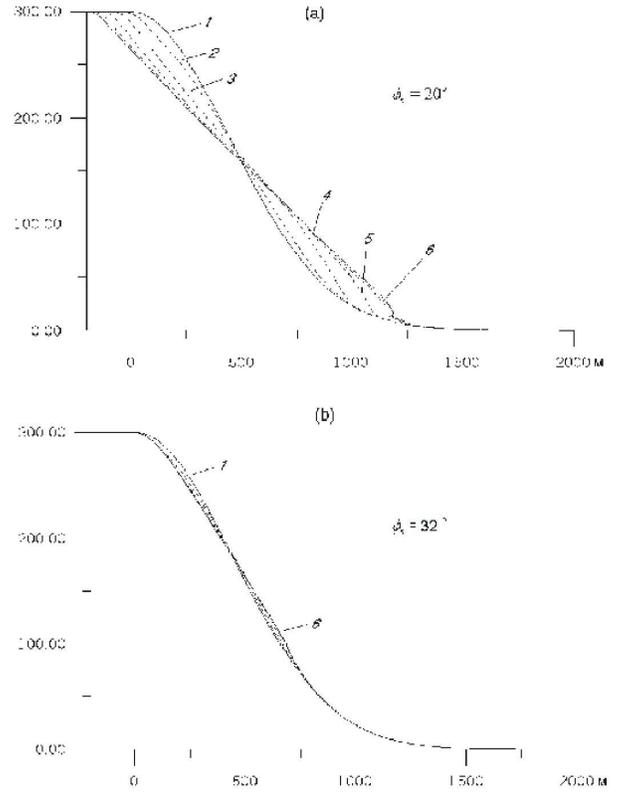}
\caption{Evolution of submarine slope surface shape during
landslide movement: a) $\phi_s = 20^o$; b) $\phi_s = 32^o$.}
\label{disp3}
\end{figure}

\begin{figure}
\includegraphics[width=8.0 cm]{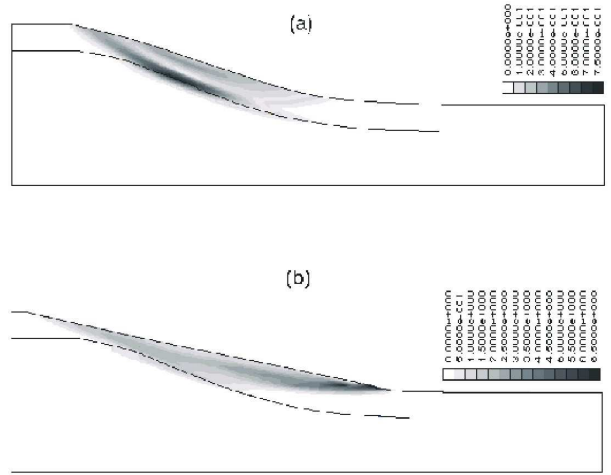}
\caption{Shear strain distribution at $\phi_s = 20^o$ a) t = 10
sec; b) t = 40 sec.}\label{disp4}
\end{figure}
In the work there are presented results of numerical simulation
for two values of maximal friction angle $\phi_s: 20^o$ and $32^o$
(see below). In fig.3a, b there are presented a picture of
successive position of landslide surface for several times with 10
sec interval (dashed curves) calculated for maximal friction angle
$\phi_s$: $20^o$ and  $32^o$, respectively. Solid line corresponds
to initial landslide shape ($t = 0$). At given values of
parameters in the case $\phi_s = 20^o$, landslide comes only
finite distance, equal to $1200 m$. The duration of sliding
process may be determined from the time instant when velocity of
forefront reaches zero, and is near 50 sec. Maximal velocity of
landslide forefront in this case is reached at near 40 sec and its
value is near $6 m/sec$. In contrast, rear part of the landslide
shifts to lower $x$. In the case $\phi_s = 32^o$, landslide comes
only near 750 m (fig. 3b) and its maximal velocity is essentially
lower though it is reached also at near 40 sec.

In fig. 4a, b there are presented a shear strain distribution in
landslide body for maximal friction angles $\phi_s = 20^o$   for
time 10 sec and 40 sec, respectively. From numerical results, it's
seen that in the case $\phi_s = 20^o$ landslide forefront reaches
level of horizontal bottom, and horizontal region near shoreline
disappears so that landslide is plane up to shoreline. In the case
of $\phi_s = 32^o$ landslide don't reach a horizontal bottom level
so that landslide shape is plane only in the middle part of slope,
and horizontal region near shoreline persists (see, \cite{GLKM}).

For numerical simulation of landslide-generated surface water
waves it was used a system of shallow water nonlinear equations
with taking into account the friction and Coriolis force $$
\frac{\partial u}{\partial t} + u\frac{\partial u}{\partial x} +
v\frac{\partial u}{\partial z} = -g\frac{\partial \eta}{\partial
x} - \frac{r}{H}u\sqrt{u^2 + v^2} + fv, \\  $$ $$ \frac{\partial
v}{\partial t} + v\frac{\partial v}{\partial z} + u\frac{\partial
v}{\partial x} = -g\frac{\partial \eta}{\partial z} -
\frac{r}{H}v\sqrt{u^2 + v^2} - fu,
\eqno(1) $$
\\[3mm]
where $x, z$ are real space coordinates, $t$ is the time, $u(x,t),
v(x,t)$ are velocity vector components, $\eta$ is the surface
elevation amplitude relative to undisturbed state, $g$ is a
gravity acceleration, $H(x, t) = h(x, t) + \eta(x, t)$ is a total
depth, ($h$ is a depth measured from undisturbed seawater level),
$r$ is a bed friction coefficient, $f$ is a Coriolis force
parameter.

Surface water waves were in fact generated by moving submarine
landslide because of continuity equation in the form

$$ \frac{\partial (h + \eta)}{\partial t} = -
\frac{\partial}{\partial x} ((h + \eta)u) -
\frac{\partial}{\partial z}((h + \eta)v) \eqno(2)
$$

To solve the system numerically it was used a scheme constructed
in analogy with those used in one-dimensional case. Numerical
simulation process was based on splitting of difference operator:
equations on $x$ and $z$ are integrated separately at two
semi-steps in time. It was used a space dispersed pattern for
different presentation of variables. Th avoid the appearance of
numerical instability problem and necessity to use a filtration
scheme it was applied a method with first order in time scheme.
The Cartesian coordinate system is taken so that total depth is
positive in the region of liquid phase and is equal to zero at the
phase interface ($\eta = - h$). So, negative values of $H$
indicate to solid phase (beach). To localize the boundary it was
used a linear extrapolation procedure. Since it is known that
upstream oriented schemes are usually stable then it was used a
conventional procedure for advective terms of equations of motion
and continuity. The procedure is based on choosing of direction of
space discretization.

In fig.5 it is presented an evolution of surface water waves
(upper curves) generated by moving submarine landslide (lower
curves) for the case of maximal friction angle $\phi_s = 20^o$.
The numbers near upper curves (surface waves) correspond to those
near lower curves (position of landslide). There arise two
distinct wave groups: crest and trough moving seaward, and second
trough moving to a beach. It is seen that wave crest velocity is
noticeably higher than that of landslide forefront. The picture is
consistent with results of numerical simulation in conventional
models (rigid body, viscous fluid, visco-plastic fluid).

\begin{figure}
\includegraphics[width=8.0 cm]{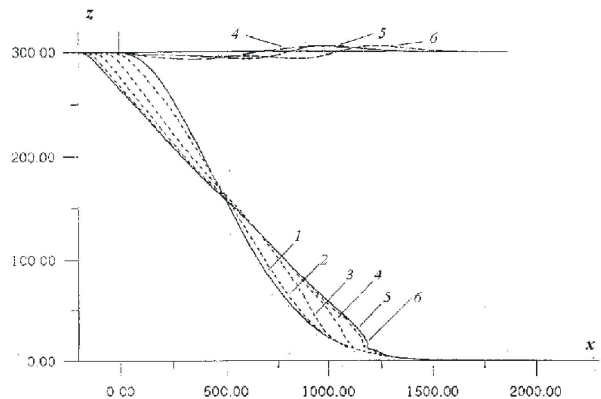}
\caption{Surface water wave generation by submarine landslide
(maximal friction angle is $\phi_s = 20^o$).}\label{disp5}
\end{figure}

In fig.6 there is presented an evolution of landslide-induced
surface water waves corresponding to maximal friction angle
$\phi_s = 20^o$ with time step 10 seconds. In the beginning, it is
generated a dipolar wave with trough oriented to a beach. This
dipolar wave moves in the deep water derection. But soon an
additional crest in the region of the trough appears which moves
to a beach. Such recession of the sea water indeed was observed in
the region where landslide occurs which phenomena leads to
anomalous tsunami with first negative phase
\cite{MP,SM}(depressive wave).
\begin{figure}
\includegraphics[width=8.0 cm]{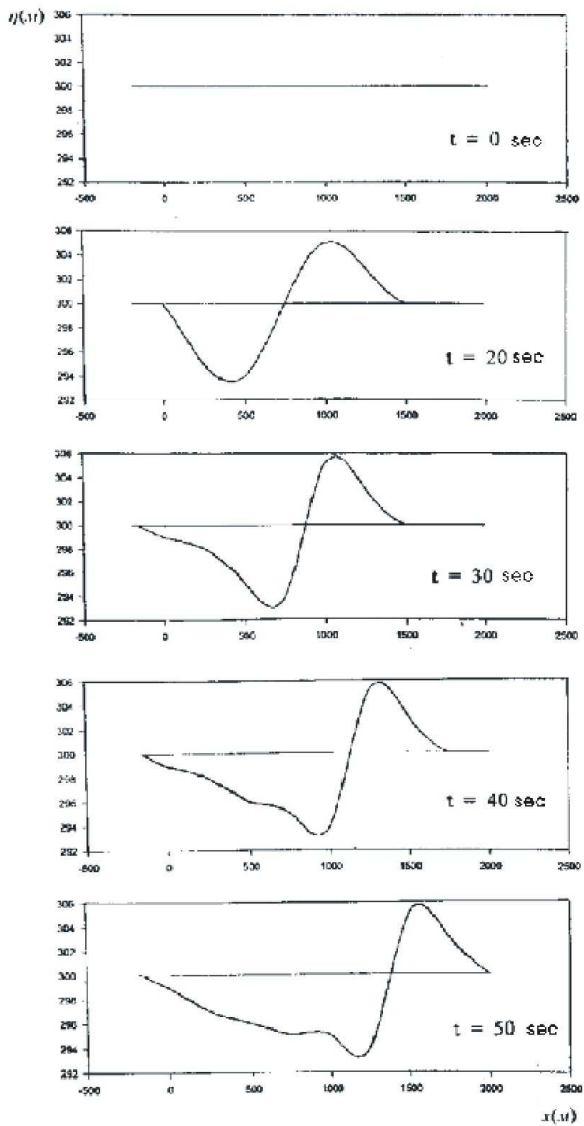}
\caption{Evolution of surface water waves generated by submarine
landslide (maximal friction angle is $\phi_s =
20^o$).}\label{disp6}
\end{figure}

In conclusion, in contrast to kinematic method, complete solution
of problem with stress strained state of slope at seismic action
gives the possibility to estimate the distribution of residual
strain and displacement in overall sediment volume with taking
into account its possible liquefaction and disconsolidation
\cite{GL}. Numerical code FLAC, in contrast to finite element
method realizes an explicit finite difference scheme for solution
of three-dimensional (3D) problems in continuum mechanics that
permits one to simulate nonlinear behavior of pore saturated mass
under conditions of plastic flow above yield stress. For
calculations, material was divided to polyhedral elements in the
limit of grid corresponding to shape of calculated object. Each
element behaves itself according to action of applied forces and
edge's restrictions. The grid is frozen in material and moves
together with it being undergone to finite strain and
displacement. Explicit Lagrange scheme of calculations guarantees
an accurate simulation and material flow. A key factor to control
the process of localization of plastic strain and related
instability is a character of constituents composing the surface
sediment layer. The essence of this phenomena is that straining of
real sediment layer after reaching of destruction point is
continued at decreasing stress, i.e. it occurs decrease of limit
of strength at increase of strain up to stress reaches some finite
or residual level. Process of sediment mass sliding is essentially
determined by friction angle. According to experiments, submarine
sediments are disconsolidated when reaching maximal strength.
Firstly, sediment mass is strained elastically but when limit
condition
$
\tau = \tau_s = c cos\phi_s - \sigma sin\phi_s
$
is fulfilled it begins to move. In result, friction angle
decreases from the peak value $\phi = \phi_s$ in the maximal
strain point, and it is established a new equilibrium state at the
level
$
\tau = \tau_k = c cos\phi_k - \sigma sin\phi_k
$
at residual value $\phi = \phi_k$ \cite{GL}. Moreover, strength
decrease of sediment mass during the development of plastic strain
at static action is a key factor to control a slope stability. In
the case of neglecting of the effect of sediment mass strength
decrease slope displacements remains to be small which fact leads
to mistake conclusions. And for further agreement of numerical
simulation results with natural data it is necessary to attract
the detailed data about inner structure of sediment formations in
landslide dangerous regions of slopes and landslide constituent
properties.

This work was supported in part by the Russian Foundation for
Basic Research (Grant No. 01-05-64548).

\end{document}